\documentclass{aa}
\input{psfig}

\def\Real{{\rm I\mathchoice{\kern-0.70mm}{\kern-0.70mm}{\kern-0.65mm}%
  {\kern-0.50mm}R}}

\font \bolditalics = cmmib10
\def\bx#1{\leavevmode\thinspace\hbox{\vrule\vtop{\vbox{\hrule\kern1pt
        \hbox{\vphantom{\tt/}\thinspace{\bf#1}\thinspace}}
      \kern1pt\hrule}\vrule}\thinspace}

\def \vc #1{{\textfont1=\bolditalics \hbox{$\bf#1$}}}

\def\eg{{\bf e}}

\def\sg{{\bf s}}
\def\dg{{\bf d}}

\def\thetag{{\vc \theta}}

\def\gammag{{\vc \gamma}}

\def\Sg{{\bf S}}

\def\Lc{{\cal L}}

\def\be{\begin{equation}}
\def\ee{\end{equation}}
\def\ba{\begin{eqnarray}}
\def\ea{\end{eqnarray}}

\def\ltsima{$\; \buildrel < \over \sim \;$}
\def\lsim{\lower.5ex\hbox{\ltsima}}
\def\gtsima{$\; \buildrel > \over \sim \;$}
\def\gsim{\lower.5ex\hbox{\gtsima}}

\begin{document}



   \title{Cosmic Shear Statistics and Cosmology \thanks{
Based on observations obtained at the
Canada-France-Hawaii Telescope (CFHT) which is operated by the National
Research Council of Canada (NRCC), the Institut des Sciences de l'Univers
(INSU) of the Centre National de la Recherche Scientifique (CNRS) and
the University of Hawaii (UH)}}

   \author{L. Van Waerbeke$^{1,2}$, Y. Mellier$^{1,3}$, M. Radovich$^{4,1}$, E. Bertin$^{1,3}$,
M. Dantel-Fort$^3$, H.J. McCracken$^5$, 
 O. Le F\`evre$^5$, S. Foucaud$^5$, J.-C. Cuillandre$^{6,7}$, T. Erben$^{1,3,8}$, B. Jain$^{9,10}$, P. Schneider$^{11}$, 
F. Bernardeau$^{12}$, B. Fort$^1$}
   \offprints{waerbeke@iap.fr}

  \institute{$^1$ Institut d'Astrophysique de Paris. 98 bis, boulevard
Arago. 75014 Paris, France. \\
$^2$ Canadian Institut for Theoretical Astrophysics, 60 St 
Georges Str., Toronto, M5S 3H8 Ontario, Canada.\\
   $^3$ Observatoire de Paris. DEMIRM. 61, avenue de
l'Observatoire.  75014 Paris, France.\\
   $^4$ Osservatorio Astronomico di Capodimonte, via Moiariello, 80131m Napoli, Italy \\
   $^5$ Laboratoire d'Astrophysique de Marseille, 13376 Marseille Cedex 12, 
France\\
   $^6$ Canada-France-Hawaii-Telescope, PO Box 1597, Kamuela, Hawaii 96743,
USA\\
   $^7$ Observatoire de Paris. 61, avenue de l'Observatoire.  75014 Paris, France.\\
   $^8$ Max Planck Institut f\"ur Astrophysiks, Karl-Schwarzschild-Str. 1,
Postfach 1523, D-85740 Garching, Germany. \\
   $^9$ Dept of Physics and Astronomy, University of Pennsylvania, 
 Philadelphia, PA 19104, USA\\
   $^{10}$ Dept. of Physics and Astronomy, Johns Hopkins University, Baltimore, MD 21218, 
USA\\
   $^{11}$ Universitaet Bonn, Auf dem H\"uegel 71, 53121 Bonn, Germany \\
   $^{12}$ Service de Physique Th\'eorique. C.E. de Saclay. 91191 Gif sur 
Yvette
Cedex, France.\\
}

   \markboth{Cosmic shear statistics}{}


   \markboth{Cosmic shear statistics}{}

\authorrunning{Van Waerbeke et al}
\abstract{We report a measurement of cosmic shear correlations
using an effective area of $6.5~sq.~deg.$ of the VIRMOS deep
imaging survey in progress at the Canada-France-Hawaii Telescope. 
We measured various shear correlation functions, the aperture
mass statistic and the top-hat smoothed variance of the shear
with a detection significance exceeding $12~\sigma$ for each
of them. We present results on angular scales from $3$
arc-seconds to half a degree. The consistency
of different statistical measures is demonstrated and confirms
the lensing origin
of the signal through tests that rely on the scalar nature of
the gravitational potential. For Cold Dark Matter models we
find $\sigma_8 \Omega_0^{0.6}=0.43^{+0.04}_{-0.05}$ at the $95\%$
confidence level. The measurement over almost three decades of scale
allows to discuss the effect of the shape of the power spectrum on the
cosmological parameter estimation.
The degeneracy on $\sigma_8-\Omega_0$ can be broken
if priors on the shape of the 
linear power spectrum (that can be parameterized by $\Gamma$) 
are assumed. For instance,
with $\Gamma=0.21$ and at the $95\%$ confidence level, we obtain
$0.6<\sigma_8<1.1$ and $0.2<\Omega_0<0.5$ for open models, and
$\sigma_8>0.65$ and $\Omega_0<0.4$ for flat ($\Lambda$-CDM) models.
From the tangential/radial modes decomposition we can set an upper limit
on the intrinsic shape alignment, which was recently suggested as a possible
contribution to the lensing signal. Within the error bars, there is
no detection of intrinsic shape alignment for scales larger than $1'$.
\keywords{Cosmology: theory, dark matter, gravitational lenses, large-scale
structure of the universe}
}

\maketitle

\section{Introduction}

Cosmological gravitational lensing produced by large-scale structure (or
cosmic shear) has been advocated as a powerful tool to probe the mass
distribution in the universe (see the reviews from
\cite{M99,BS01} and references therein).  The first detections  
  reported  over the past year
(\cite{VW00,BRE00,K00,WT00,Ma01,RRG01}) confirmed that the amplitude and the 
 shape of the signal are compatible with theoretical expectations, although 
the data sets were not 
 large enough to place useful constraints
on cosmological models.  \cite{Ma01} combined the results from different
groups to obtain constraints on the power spectrum normalization
$\sigma_8$ and the mean density of the universe $\Omega_0$:
Their result is in agreement with the cluster abundance 
constraints, but they were not yet able to break the degeneracy 
between $\sigma_8$ and $\Omega_0$.

The physical interpretation of the weak lensing signal can
be made more securely using detections of cosmic
shear from different statistics and angular scales 
on the {\it same} data set (as in \cite{VW00}).
Unfortunately, their joint detection of the variance and the correlation
function using the same data was not fully conclusive:
the sample was too small to enable a significant 
detection of the cosmic shear from
variances with different weighting schemes
and $2$-points statistics over a wide range of scales.
The use of independent approaches is nevertheless necessary and it is
a crucial step to validate the reliability of cosmic shear, to check the
consistency of the measurements against theoretical
predictions and to understand the residual systematics. 
A relevant example is the
aperture mass statistic (defined in \cite{SVWJK98}). It is a direct
probe of the projected mass power spectrum, and it is not sensitive to
certain type of systematics (like a uniform PSF anisotropy)
which may corrupt the top-hat smoothed variance, or
the shear correlation function. Even the shear correlation function 
can be measured in
several ways, by splitting the tangential and radial modes for instance. 

In this paper we report the measurement of the top-hat smoothed
variance, the aperture mass, the shear correlation function,
and the tangential and radial shear correlation functions
on a new homogeneous data set covering an effective area of
6.5 square-degrees (deg$^2$). The depth 
and the field of view are well suited for a comprehensive
analysis using various statistics. 
We show that the amplitude of residual systematics is very low 
compared to the signal and discuss the consistency of
these measurements against the predictions of cosmological models. 

We also discuss alternative interpretations.  
It has been suggested recently that intrinsic alignments of
galaxies caused by tidal fields could contribute to the lensing
signal (\cite{CM00,HRH00,CKB00,CNPT00a,CNPT00b}). This type
of systematic is problematic because its signature on
different $2$-points statistics mimics the lensing effect.
A mode decomposition in {\it electric} and {\it magnetic}
types (or $E$ and $B$ modes), similar to what is performed for
the polarization analysis in the Cosmic Microwave Background,
can separate lensing from intrinsic alignment
(see \cite{CNPT00a,CNPT00b}). The $E$ and $B$ mode analysis is
the subject of a forthcoming paper; 
the aperture mass statistic presented in this paper is a similar
analysis to the $E$ and $B$ mode decomposition, and allows us
to put an upper limit on the
contamination of our survey by the intrinsic alignments.

This paper is organized as follow: Section 2 describes
our data set, and highlights the differences in the data preprocessing
from our previous analysis (\cite{VW00}). The measurement of the shear
from this imaging data is discussed in Section 3. Section 4 summarizes the
theoretical aspects of the different quantities we measure, and lists
the statistical estimators used. The results and comparison to a few standard
cosmological models are shown in Section 5. In Section 6 we perform a
maximum likelihood analysis of cosmological models in the $(\Omega_0,
\sigma_8)$ parameter space. The results on very small scales are shown 
separately in Section 7, and we conclude in Section 8.

\section{The data set}
The DESCART weak lensing project\footnote{http://terapix.iap.fr/Descart} 
is a theoretical and observational program for cosmological 
  weak lensing investigations.   The cosmic shear survey carried out 
by the DESCART team  uses the CFH12K data jointly with the VIRMOS 
survey\footnote{http://www.astrsp-mrs.fr/virmos/} to  produce a 
large homogeneous 
  photometric sample which will eventually contain a catalog of galaxies with 
redshifts as well as the projected mass density over the whole 
field (\cite{olf01}).  
  In contrast to \cite{VW00}, the 
  new sample presented in this work only uses I-band data taken with 
the CFH12K camera and is therefore more homogeneous.
It is worth noting that our new CFH12K sample only uses half of the 
data of the previous one.  A comparison of the results will also permit
to check the consistency and the robustness of the cosmic shear analysis.

The CFH12K data was obtained during dark nights 
  in May 1999, November 1999 and April 2000 following 
  the standard  observation procedure
  described in Van Waerbeke et al (2000).
  The fields are spread over 4 independent $2\times2$ deg$^2$ 
   areas of the sky identified as F02, F10, F14 and F22. Each 
  field is a compact mosaic of 16 CFH12K pointings named P[n]$_{n=1-16}$.  
Once  the survey is completed  each of them will cover 4 deg$^2$. Currently,   
 of the final 16 deg$^2$, only 8.38 deg$^2$ is available for 
the analysis -- most of 
the pointings are  located in three different fields (F02, F10, F14 listed
in Table \ref{targets}). This total 
  field of view gets significantly reduced by the masking and
selection  procedures described below.  A summary of 
  the data set characteristics are listed in Table  \ref{targets}.

\begin{table*}
\caption{List of the fields. All observations were done in I band
 with the CFH12K camera (\cite{Cuill00}). The number following the F denotes the field
name, and the number following the P denotes the pointing name within the
field. The geometry of the survey is detailed in 
http://terapix.iap.fr/Descart/. The image quality has been measured on
   each stacked image from a standard fitting of a Moffat profile.
}
\label{targets}
\begin{center}
\begin{tabular}{|c|c|c|c|c|c|c|c|}
\hline
Target & Used area &  Exp. time & Period & Image quality\\
\hline
F02P1 &  980 ${\rm arcmin}^2 $ & 9390 sec. & Nov. 1999 & 0.75" \\
F02P2 &  1078 ${\rm arcmin}^2 $ & 7200 sec. & Nov. 1999 & 0.90" \\
F02P3 &  980 ${\rm arcmin}^2 $ & 7200 sec. & Nov. 1999 & 0.90" \\
F02P4 &  1078 ${\rm arcmin}^2 $ & 7200 sec. & Nov. 1999 & 0.80" \\
F10P1 &  882 ${\rm arcmin}^2 $ & 3600 sec. & May 1999 & 0.65" \\
F10P2 &  882 ${\rm arcmin}^2 $ & 3600 sec. & May 1999 & 0.75" \\
F10P3 &  490 ${\rm arcmin}^2 $ & 3600 sec. & May 1999 & 0.75" \\
F10P4 &  882 ${\rm arcmin}^2 $ & 3600 sec. & May 1999 & 0.65" \\
F10P5 &  882 ${\rm arcmin}^2 $ & 3600 sec. & May 1999 & 0.75" \\
F10P7 &  1176 ${\rm arcmin}^2 $ & 3600 sec. & Apr. 2000 & 0.75" \\
F10P8 &  1176 ${\rm arcmin}^2 $ & 3600 sec. & Apr. 2000 & 0.70" \\
F10P9 &  98 ${\rm arcmin}^2 $ & 3600 sec. & Apr. 2000 & 0.65" \\
F10P10 &  784 ${\rm arcmin}^2 $ & 3600 sec. & Nov. 1999 & 0.80" \\
F10P11 &  294 ${\rm arcmin}^2 $ & 3600 sec. & Nov. 1999/Apr. 2000 & 0.90" \\
F10P12 &  1176 ${\rm arcmin}^2 $ & 3600 sec. & Apr. 2000 & 0.80" \\
F10P15 &  686 ${\rm arcmin}^2 $ & 3600 sec. & Apr. 2000 & 0.85" \\
F14P1 &  882 ${\rm arcmin}^2 $ & 3600 sec. & May 1999 & 0.80" \\
F14P2 &  882 ${\rm arcmin}^2 $ & 3600 sec. & May 1999 & 0.85" \\
F14P3 &  686 ${\rm arcmin}^2 $ & 3600 sec. & May 1999 & 0.75" \\
F14P4 &  1078 ${\rm arcmin}^2 $ & 3600 sec. & May 1999 & 0.75" \\
F14P5 &  980 ${\rm arcmin}^2 $ & 3600 sec. & May 1999 & 0.70" \\
F14P6 &  686 ${\rm arcmin}^2 $ & 3600 sec. & May 1999 & 0.80" \\
F14P7 &  686 ${\rm arcmin}^2 $ & 3600 sec. & May 1999 & 0.70" \\
F14P8 &  882 ${\rm arcmin}^2 $ & 3600 sec. & May 1999 & 0.85" \\
F14P9 &  1078 ${\rm arcmin}^2 $ & 3600 sec. & Apr. 2000 & 0.75" \\
F14P10 & 784 ${\rm arcmin}^2 $ & 3600 sec. & May 1999 & 0.85" \\
F14P11 & 882 ${\rm arcmin}^2 $ & 3600 sec. & Apr. 2000 & 0.80" \\
F14P12 & 784 ${\rm arcmin}^2 $ & 3600 sec. & Apr. 2000 & 0.80" \\
F14P13 & 882 ${\rm arcmin}^2 $ & 3600 sec. & Apr. 2000 & 0.85" \\
F14P14 & 882 ${\rm arcmin}^2 $ & 3600 sec. & May 1999 & 1.0" \\
F14P15 & 882 ${\rm arcmin}^2 $ & 3600 sec. & Apr. 2000 & 0.90" \\
F14P16 & 1176 ${\rm arcmin}^2 $ & 2880 sec. & Apr. 2000 & 0.65" \\
F22P3 & 686 ${\rm arcmin}^2 $ & 3600 sec. & May 1999 & 0.75" \\
F22P4 & 980 ${\rm arcmin}^2 $ & 3600 sec. & Nov. 1999 & 0.75" \\
F22P6 & 588 ${\rm arcmin}^2 $ & 3600 sec. & Apr. 2000 & 0.80" \\
F22P11 & 294 ${\rm arcmin}^2 $ & 2880 sec. & Apr. 2000 & 0.75" \\
\hline
\end{tabular}
\end{center}
\end{table*}

The data reduction was done at the TERAPIX data
 center\footnote{http://terapix.iap.fr}.  More than 
  1.5 Tbytes of data were processed in order to 
   produce the final stacked images. The reduction procedure is the same as 
 in \cite{VW00}, so we refer to this paper for the details.  
However, in order to improve the image quality prior to correction for the PSF 
anisotropy and to get a better signal-to-noise ratio on a larger angular scale 
 than in our previous work,  all CFH12K images were co-added after 
 astrometric corrections. 

The astrometric calibration and the co-addition were done using the MSCRED 
package in IRAF. Some tasks have been modified in order to allow a fully
automatic usage of the package. For each pointing, we first started with the
images in the I band. An astrometric solution was first found for one set of
exposures in the dither sequence using the USNO-A 2.0 as reference, which 
provides the position of $\sim \times10^8$ sources with an RMS accuracy of 
0.3 arcsec. The astrometric solution was then transferred to the other
exposures in the sequence. All object catalogs were  obtained using
SExtractor (\cite{BA96})\footnote{ftp://ftp.iap.fr/pub/from\_users/bertin/sextractor/}
   and a linear correction to the world coordinate system was computed
with respect to the initial set. Finally, all images were resampled using 
a bi-cubic interpolation and then stacked together. 

At this stage, each stacked image was inspected by eye and 
   all areas which may potentially influence the later lensing analysis
signal were masked (see \cite{VW00} and \cite{Ma01}).  Since we 
   adopted conservative masks, this process had a dramatic impact 
  on the field of view: we lost 20\% of the total area and 
  ended up with a usable area of 6.5 deg$^2$.

 The photometric calibrations were done using standard stars from the 
  Landolt's catalog (\cite{landolt}) covering a broad sample of 
  magnitude and colors. A full description of the photometric procedure 
 is beyond the scope of this work and will be discussed elsewhere 
(Le F\`evre et al, in preparation). In summary, we used the SA110 and SA101 
  star fields to measure the zero-points and color equations of each run.
  From these calibrations, we produced the magnitude histograms of each
  field in order to find out the cut off and a rough limiting magnitude.
 Although few fields have exposure time significantly larger than 
  1 hour, the depth of the sample is reasonably stable from field 
 to field and 
reaches $I_{AB}=24.5$. Up to this magnitude, 1.2 million galaxies were 
detected over the total area of 8.4 deg$^2$.
  
\section{Shear measurements}

\subsection{Shape measurement}

The details of our shape measurement procedure and Point Spread
Function (hereafter PSF) correction have been extensively described in two
previous papers (\cite{VW00,Ma01}), and tested against numerical
simulations (\cite{E00}). Therefore we will not
reproduce these details here, but only give a short overview of
the procedure. The shape
measurement pipeline uses the IMCAT software (\cite{KSB95})\footnote{kindly
made available by Nick
Kaiser at http://www.ifa.hawaii.edu/$\sim$kaiser/} combined with the SExtractor
 package.
The different steps in the procedure are as follows:

\begin{itemize}
\item Object detection with Sextractor.
\item The shape parameters defined in \cite{KSB95} are calculated
using IMCAT.
\item Stars are identified in the stellar branch of the size-magnitude
diagram. Stars brighter than 1 magnitude below the saturation
level are excluded. Objects smaller than the PSF size are
discarded as galaxy candidates (because a shape correction
below the PSF size is meaningless).
\item The PSF is measured from the stars, and interpolated
continuously over the CCD's using a third order polynomial.
\item Galaxy shapes are corrected using the scheme developed
in \cite{KSB95}, modified and adapted to our problem as described
in \cite{E00}. The shape correction is a two-step process: first
we remove the anisotropic contribution of the PSF, then the isotropic
contribution is suppressed according to \cite{LK97}.
\item A weight $w$ is calculated for each galaxy, which depends on
the level of noise in the shape correction (see Eq.(7) in \cite{VW00}).
\item For each galaxy pair with members closer than 15 pixels (3
arcsec), one member is removed, in order to avoid the problem of 
overlapping isophotes reported in \cite{VW00}.
\item Each CCD is visualized by eye, and the bad areas are masked
(star spikes and ghost images, blank lines or columns, fringe
residuals). After the whole process of cleaning and object
selection, 420,000 galaxies were effectively used for the
weak lensing analysis.
\end{itemize}

\begin{figure}
\centerline{
\psfig{figure=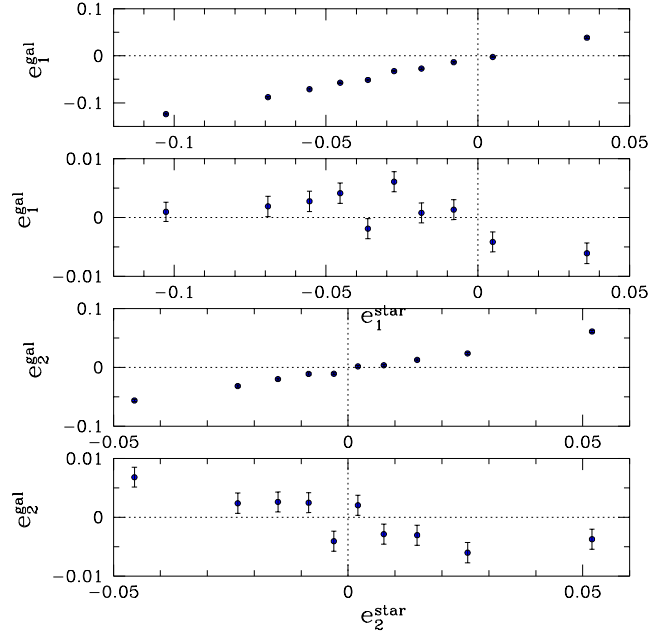,height=9cm}}
\caption{\label{systematics.ps} Top and third from the top
panels: averaged $e^{\rm gal}_1$ and $e^{\rm gal}_2$
component versus the ellipticity of the PSF at the galaxy location,
{\it without} the anisotropic PSF correction. Second from top and bottom
panels: averaged $e^{\rm gal}_1$ and $e^{\rm gal}_2$
{\it including} the anisotropic PSF correction.
}
\end{figure}

The raw ellipticity
$\eg$ of a galaxy is measured from the second moments $I_{ij}$ of the surface
brightness $f(\thetag)$:

\begin{equation}
\eg=\left({I_{11}-I_{22}\over Tr(I)} ; {2I_{12}\over Tr(I)}\right), \ \ \
I_{ij}=\int {\rm d}^2\theta
W(\theta)\theta_i\theta_j f(\thetag).
\end{equation}
The window function $W(\thetag)$ suppresses the noise at large distances
from the object center. The procedure described above gives a corrected
galaxy ellipticity $\eg^{\rm gal}$ calculated from the $\eg$'s.
According to \cite{KSB95}, the ensemble
average of  $\eg^{\rm gal}$ is equal to the shear $\gammag$ at the galaxy
location.
Figure \ref{systematics.ps} shows the level of systematics in $\eg^{\rm gal}$
with and without the anisotropic PSF correction. After the PSF
correction, the average galaxy
ellipticity is bounded between $-0.005$ and $0.005$, and the variance
of the residual systematics is less than $\sim 10^{-5}$. As
we shall see later this is
much less than the measured signal. As quoted in \cite{VW00}, the
galaxy ellipticities show a small offset $(-0.008, -0.003)$, which has been
corrected in this figure (the origin of this offset is still
unclear). However it is worth to mention that the aperture mass is not sensitive
to this offset.

\section{Statistical measures of shear correlations}

\subsection{Theory}

We summarize the different statistics we shall measure, and
how they depend on cosmological models.
We concentrate on 2-points statistics and variances, since
higher order moments are more difficult to measure, and will be 
 addressed in a forthcoming paper.

Let us assume a source redshift distribution parameterized as:

\begin{equation}
n(z_s)={\beta\over z_0 \ \Gamma\left({1+\alpha\over \beta}\right)} 
\left({z_s\over
z_0}\right)^\alpha \exp\left[-\left({z_s\over z_0}\right)^\beta\right],
\label{zsource}
\end{equation}
where the parameters $(z_0,\alpha,\beta)=(0.8,2,1.5)$, which is consistent
with a limiting magnitude $I_{AB}=24.5$ given by \cite{C00} (it corresponds
to a mean redshift of $1.2$).
We define the power spectrum of the convergence as 
 (following the notation in \cite{SVWJK98}):

\begin{eqnarray}
P_\kappa(k)&=&{9\over 4}\Omega_0^2\int_0^{w_H} {{\rm d}w \over a^2(w)}
P_{3D}\left({k\over f_K(w)};
w\right)\times\nonumber\\
&&\int_w^{w_H}{\rm d} w' n(w') {f_K(w'-w)\over f_K(w')},
\label{pofkappa}
\end{eqnarray}
where $f_K(w)$ is the comoving angular diameter distance out to a distance $w$
($w_H$ is the horizon distance),
and $n(w(z))$ is the redshift distribution of the sources given in
Eq.(\ref{zsource}). $P_{3D}(k)$ is the non-linear mass power spectrum,
and $k$ is the 2-dimensional wave vector perpendicular to the
line-of-sight.
For a top-hat smoothing window of radius $\theta_c$, the
variance is:

\begin{equation}
\langle\gamma^2\rangle={2\over \pi\theta_c^2} \int_0^\infty~{{\rm d}k\over k} P_\kappa(k)
[J_1(k\theta_c)]^2,
\label{theovariance}
\end{equation}
where $J_1$ is the first Bessel function of the first kind.\\
The aperture mass $M_{\rm ap}$ was introduced in \cite{K94}:

\begin{equation}
M_{\rm ap}=\int_{\theta < \theta_c}~{\rm d}^2\thetag \kappa(\thetag)~U(\theta),
\end{equation}
where $\kappa(\thetag)$ is the convergence field, and $U(\theta)$ is a
compensated filter
(i.e. with zero mean). \cite{SVWJK98} applied this statistic to the cosmic
shear measurements. They showed that the aperture mass variance is related to
the convergence power spectrum by:

\begin{equation}
\langle M_{\rm ap}^2\rangle={288\over \pi\theta_c^4} \int_0^\infty~{{\rm d}k\over k^3}
 P_\kappa(k) [J_4(k\theta_c)]^2.
\label{theomap}
\end{equation}
$\langle M_{\rm ap}^2\rangle$ can be calculated directly from the shear
$\gammag$ without the need for a mass reconstruction. \\
For each galaxy, we
define the tangential and radial shear components ($\gamma_t$ and $\gamma_r$)
with respect to the center of the aperture:

\begin{eqnarray}
\gamma_t &=& -\gamma_1 \cos(2\phi)-\gamma_2 \sin(2\phi) \nonumber\\
\gamma_r &=& -\gamma_2 \cos(2\phi)+\gamma_1 \sin(2\phi),
\label{modedef}
\end{eqnarray}
where $\phi$ is the position angle between the x-axis and the line connecting
the aperture center to the galaxy.
It is then easy to show that the aperture mass is related to the tangential shear
by:

\begin{equation}
M_{\rm ap}=\int_{\theta < \theta_c}~{\rm d}^2\thetag \gamma_t(\thetag)~Q(\theta),
\label{mapfromshear}
\end{equation}
where the filter $Q(\theta)$ is given from $U(\theta)$:

\begin{equation}
Q(\theta)={2\over \theta^2}\int_0^\theta~{\rm d}\theta'~\theta'~U(\theta')-U(\theta)
\label{Qfct}
\end{equation}
If $\gamma_t$ is replaced by $\gamma_r$ in Eq.(\ref{mapfromshear}), then
the lensing signal vanishes, due to the curl-free property
of the shear field (\cite{K94})\footnote{Curl modes are produced
by non-linear lensing effects, but these are very small (\cite{B97}).}.
This remarkable property constitutes a test of the lensing origin of the
signal. The change from $\gamma_t$ to $\gamma_r$ can simply 
be accomplished just by rotating the galaxies by $45$ degrees
in the aperture (i.e. changing a curl-free field to a pure curl field).
Hereafter we call the $M_{\rm ap}$ statistic measured with the
45 degrees rotated galaxies the $R$-mode ($R$ for radial mode), and
$\langle M_\perp^2\rangle$ the corresponding variance.
It is interesting to note that the $R$-mode is not expected to vanish
if the measured signal is due to intrinsic alignments of galaxies
(\cite{CNPT00b}). Therefore it can be
used to constrain the amount of residual systematics as well as the
degree of the intrinsic alignment of galaxies.
\\
From the shear $\gammag$ and its projections defined in 
Eq.(\ref{modedef}) we can also define
various galaxy pairwise correlation functions related to the
convergence power spectrum.
Note that the tangential and radial shear projections in what follows 
are performed using the relative location vector of the 
pair members, not from an aperture center.
The following correlation functions can be defined (\cite{ME91,K92}):

\begin{equation}
\langle\gamma\gamma\rangle_\theta={1\over 2\pi} \int_0^\infty~{\rm d} k~
 P_\kappa(k) J_0(k\theta),
\label{theogg}
\end{equation}
\begin{equation}
\langle\gamma_t\gamma_t\rangle_\theta={1\over 2\pi} \int_0^\infty~{\rm d} k~
 P_\kappa(k) [J_0(k\theta)+J_4(k\theta)],
\label{theoetet}
\end{equation}
\begin{equation}
\langle\gamma_r\gamma_r\rangle_\theta={1\over 2\pi} \int_0^\infty~{\rm d} k~
 P_\kappa(k) [J_0(k\theta)-J_4(k\theta)],
\label{theoerer}
\end{equation}
where $\theta$ is the pair separation angle. The cross-correlation
$\langle\gamma_t\gamma_r\rangle_\theta$ is expected to vanish for
parity reasons (there is no preferred orientation on average).

It is easy to see that the
Eqs.(\ref{theovariance},\ref{theomap},\ref{theogg},\ref{theoetet},\ref{theoerer})
are different
ways to measure the {\it same} quantity, that is the convergence power
spectrum $P_\kappa(k)$.
Ultimately the goal is to deproject $P_\kappa(k)$ in order to
reconstruct the 3D mass power spectrum from Eq.(\ref{pofkappa}),
but this is beyond the scope of
this paper. Here we restrict our analysis to a joint detection
of these statistics, and show that they are consistent with
the gravitational lensing hypothesis. We will also examine the
constraints on the power spectrum normalization $\sigma_8$ and the
mean density of the universe $\Omega_0$.

\subsection{Estimators}

Let us now define the estimators we used to measure the quantities given in
Eqs.(\ref{theovariance},\ref{theomap},\ref{theogg},\ref{theoetet},\ref{theoerer}).

The variance of the shear is simply obtained by a cell averaging
of the squared shear $\gamma^2(\thetag_i)$ over the cell index $i$. An
unbiased estimate of the squared shear for the cell $i$ is:

\begin{equation}
E[\gamma^2(\thetag_i)]={\displaystyle \sum_{\alpha=1}^2 \sum_{k\ne l}^{N_i} w_k w_l
~e^{\rm gal}_\alpha(\vec\theta_k) e^{\rm gal}_\alpha(\vec\theta_l)
\over
\displaystyle \sum_{k\ne l}^{N_i} w_k w_l},
\label{variance}
\end{equation}
where $w_k$ is the weight for the galaxy $k$, and $N_i$ is the number of galaxies
in the cell $i$. The cell averaging over the survey is then an unbiased
estimate of the shear variance $\langle \gamma^2\rangle$. However, due to
the presence of masked areas (mentioned in Section 4.1),
some cells may have a very
low number of galaxies compared to others. Instead of applying an arbitrary sharp
cut off on the fraction of the apertures filled with masks
(as it was in previous works) we decided to keep all the cells, and to weight
each of them with the squared sum of the galaxy weights located in the cell.
The cell averaging is now defined as:

\begin{equation}
E[\gamma^2]={\displaystyle \sum_{\rm cells}
\left[E[\gamma^2(\thetag_i)] \left(\sum_{k=1}^{N_i} w_k\right)^2\right]
\over \displaystyle \sum_{\rm cells} \left[\left(\sum_{k=1}^{N_i} w_k\right)^2\right]},
\label{variance_field}
\end{equation}
where $i$ identifies the cell. One potential problem with this
procedure is that
the sum of the weights is related to the number of objects in
the aperture, which is affected by magnification bias, and therefore
correlated with the shear signal measured in the same aperture. Fortunately
the first non-vanishing contribution of this weighting scheme
is a third order effect (of order $1\%$), and is therefore 
negligible\footnote{Moreover
the slope of number counts in our I-band is $\sim 0.3$, which makes the
magnification effect very small (see \cite{M98} for an application
of the effect to the angular correlation function).}.
The advantage is that we can use {\it all}  
cells without wondering about their filling factor, and it
naturaly down-weights the cells which contain a large fraction of
poorly determined galaxy ellipticities. The weighting scheme
Eq.(\ref{variance_field}) has been tested against numerical simulation, using
a simulated survey with exactly the same survey geometry as our data: it
gave unbiased measures of the lensing signal applied to the galaxies.

The $M_{\rm ap}$ statistic is calculated from a similar
estimator, although the smoothing window is no longer a top-hat but
the $Q$ function defined in Eq.(\ref{Qfct}). An unbiased estimate of
$M_{\rm ap}^2(\thetag_i)$ in the cell $i$ is:

\begin{equation}
E[M_{\rm ap}^2(\thetag_i)]={\displaystyle \sum_{k\ne l}^{N_i} w_k w_l
~e^{\rm gal}_t(\vec\theta_k) e^{\rm gal}_t(\vec\theta_l)
~Q(\theta_k)Q(\theta_l)
\over
\displaystyle \sum_{k\ne l}^N w_k w_l},
\label{map}
\end{equation}
where $e_t^{\rm gal}$ is the tangential galaxy ellipticity, and $Q$ is given
by (see \cite{SVWJK98}):

\begin{equation}
Q(\theta)={6\over \pi}\left({\theta\over \theta_c}\right)^2
\left[1-\left({\theta\over \theta_c}\right)^2\right].
\end{equation}
The estimation of $\langle M_{\rm ap}^2\rangle$ over the survey is then
given by the same expression 
as in Eq.(\ref{variance_field}), with $E[\gamma^2(\thetag_i)]$
replaced by $E[M_{\rm ap}^2(\thetag_i)]$. We emphasize that the
this filter probes
effective scales $\theta_c/5$, and not $\theta_c$ (see Figure
2 in \cite{SVWJK98}). Therefore we have to be careful when comparing the
signal at different scales between different estimators.

The shear correlation function $\langle\gamma\gamma\rangle_\theta$
at separation $\theta$ is obtained by identifying all the pairs of galaxies
falling in the separation interval $[\theta-d\theta,\theta+d\theta]$,
and calculating the pairwise shear correlation:

\begin{equation}
E[\gamma\gamma;\theta]={\displaystyle \sum_{\alpha=1}^2}{\displaystyle
\sum_{pairs} w_k w_l
~e^{\rm gal}_\alpha(\vec\theta_k) e^{\rm gal}_\alpha(\vec\theta_l)
\over \displaystyle \sum_{pairs} w_k w_l}.
\label{shearcorrel}
\end{equation}
The tangential and radial correlation functions
$\langle\gamma_t\gamma_t\rangle_\theta$ and
$\langle\gamma_r\gamma_r\rangle_\theta$
are measured also from Eq.(\ref{shearcorrel}) by replacing
$\eg^{\rm gal}$ with $e_t^{\rm gal}$ and $e_r^{\rm gal}$ respectively
and dropping the sum over $\alpha$.
It is worth noting that the estimators given here are
independent of the angular correlation properties of the
source galaxies. 

\section{Results and comparison to cosmological models}

\begin{figure}
\centerline{
\psfig{figure=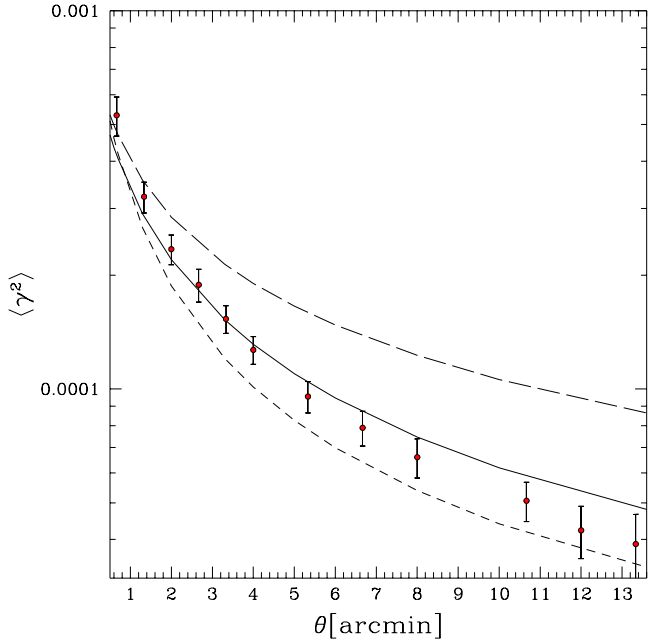,height=9cm}}
\caption{\label{tophat.ps} Top-hat smoothed variance of the shear (points
with error bars). The three models correspond to
$(\Omega_0,\Lambda,\sigma_8)=(0.3,0,0.9),(0.3,0.7,0.9),(1,0,0.6)$ for
the short-dashed, solid and long-dashed lines respectively. The power spectrum
is a CDM-model with $\Gamma=0.21$. The error bars correspond to the dispersion
of the variance measured from $200$ realizations of the data set
with randomized orientations of the galaxy ellipticities. 
}
\end{figure}
\begin{figure}
\centerline{
\psfig{figure=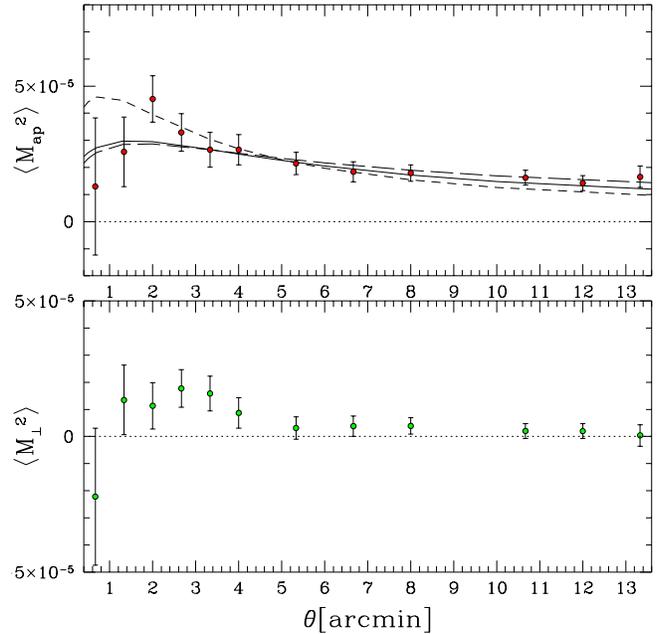,height=9cm}}
\caption{\label{map.ps} The aperture mass statistic for the same  
models as in Figure \ref{tophat.ps}. The lower panel plots the 
R-mode, obtained by making a 45 degree rotation as described in the text. 
There is no significant detection for $\theta > 5~{\rm arcmin}$ (corresponding
to an effective angular scale of $1'$, as discussed in the text), which
shows the low level of contamination by galaxy intrinsic 
alignment and/or residual systematics.
}
\end{figure}
\begin{figure}
\centerline{
\psfig{figure=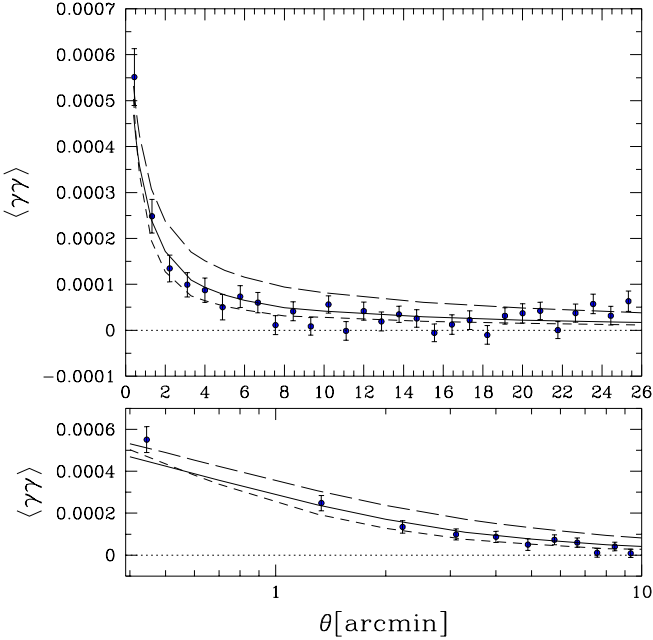,height=9cm}}
\caption{\label{gg.ps} Shear correlation function
$\langle \gamma\gamma\rangle_\theta$. The models are the same as in Figure
\ref{tophat.ps}. The lower panel uses a log-scale for the x-axis to
highlight the small scale details.
}
\end{figure}

\begin{figure}
\centerline{
\psfig{figure=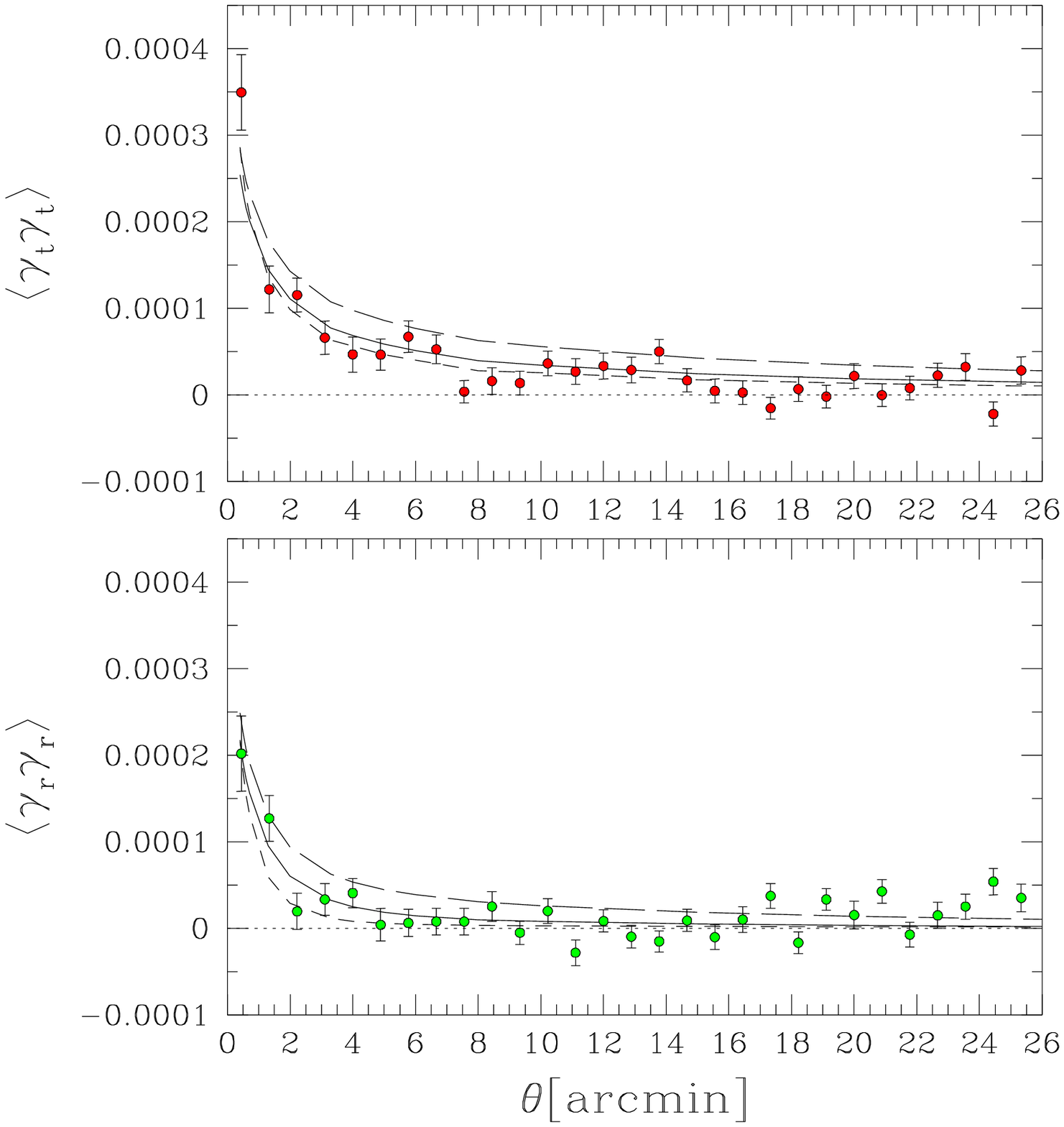,height=9cm}}
\caption{\label{eteterer.ps} Top panel: tangential shear correlation function
$\langle \gamma_t\gamma_t\rangle_\theta$. Bottom panel: radial shear
correlation function $\langle \gamma_r\gamma_r\rangle_\theta$.
The models are the same as in Figure \ref{tophat.ps}.
}
\end{figure}

\begin{figure}
\centerline{
\psfig{figure=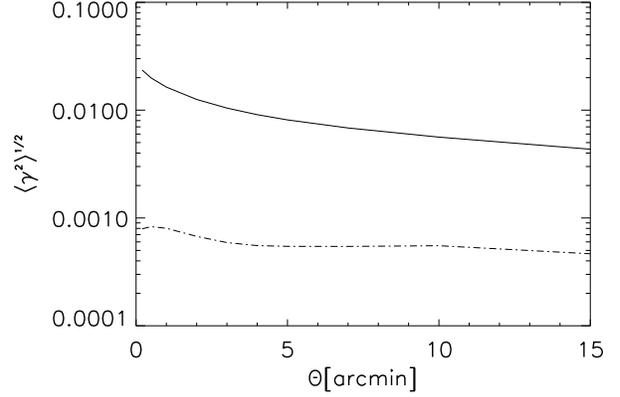,height=6cm}
}
\caption{\label{cosmicvar.ps} Shear r.m.s. $\langle \gamma^2\rangle^{1/2}$
(solid line) measured in a ray-tracing simulation (\cite{JSW00}) for the
open $\Omega_0=0.3$ model. The dashed line is the sample variance of
the shear r.m.s. measured
from 7 different realisations of the mass distribution for a survey of
$6.5$ sq. degrees.
}
\end{figure}

\begin{figure}
\centerline{
\psfig{figure=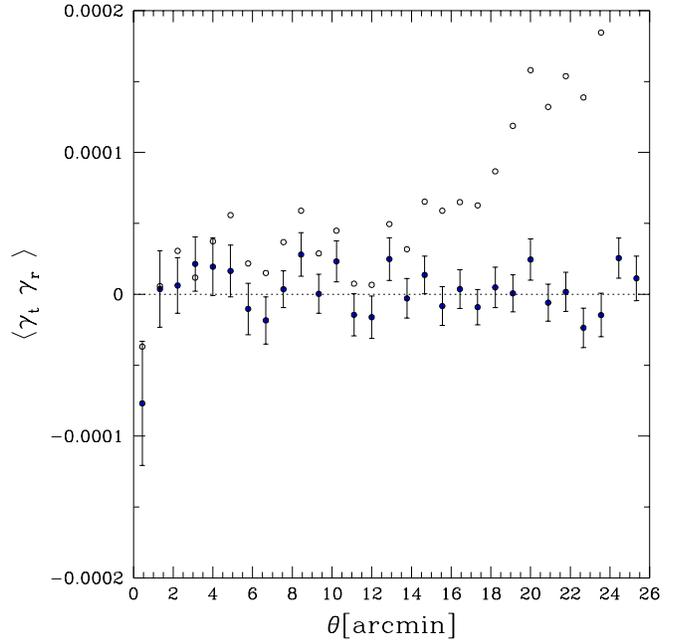,height=9cm}}
\caption{\label{crosscorrel.ps} Shear cross-correlation function
$\langle \gamma_t\gamma_r\rangle_\theta$. The signal should vanish
if the data are not contaminated by systematics. As a comparison, the
open circles show the same cross-correlation function computed from the
galaxy ellipticities where the anisotropic correction of the PSF has been
skipped. 
}
\end{figure}

In this section we present our measurements of the 
2-point correlations of the shear using the different estimators
defined above. Figures \ref{tophat.ps} to \ref{crosscorrel.ps} show the 
results for the different estimators: the variance in Figure \ref{tophat.ps}, 
the mass aperture statistic in Figure \ref{map.ps}, the shear 
correlation function in Figure \ref{gg.ps}, the radial and
tangential shear correlations in Figure \ref{eteterer.ps}, and the
cross-correlation of the radial and tangential shear in Figure 
\ref{crosscorrel.ps}. Along with the measurements we
show the predictions of three cosmological models which are 
representative of an open model, a flat model with cosmological 
constant, and an Einstein-de Sitter model. The amplitude of mass 
fluctuations in these models is 
normalized to the abundance of galaxy clusters. The three models are
characterized by the values of $\Omega_0, \Lambda$ and $\sigma_8$ as 
follows: 

\begin{itemize}
\item short-dashed line: $\Omega_0=0.3$, $\Lambda=0$, $\sigma_8=0.9$
\item solid line: $\Omega_0=0.3$, $\Lambda=0.7$, $\sigma_8=0.9$
\item long-dashed line: $\Omega_0=1$, $\Lambda=0$, $\sigma_8=0.6$
\end{itemize}
The power spectrum is taken to be a cold dark matter (CDM) power spectrum
with shape parameter $\Gamma=0.21$. The predictions for shear correlations
are computed using the non-linear evolution of the power spectrum
using the \cite{PD96} fitting formula. The source redshift
distribution follows Eq.(\ref{zsource}) with 
$(z_0,\alpha,\beta)=(0.8,2,1.5)$, which corresponds to a mean redshift of 1.2. 

It is reassuring that the different statistics agree with each other 
in their comparison with the model predictions. These
statistics weight the data in different ways and are susceptible to 
different kinds of systematic errors. 
The consistency of all the 2-point estimators suggests that
the level of systematics in the data is
low compared to the signal. A further test for systematics is
provided by measuring the cross-correlation function 
$\langle \gamma_t\gamma_r\rangle_\theta$, which should be zero for
a signal due to gravitational lensing. Figure \ref{crosscorrel.ps}
shows the measured cross-correlation function, which is indeed consistent
with zero on all scales. The figure also shows the 
cross-correlation obtained when the anisotropic contamination of the
PSF is not corrected -- clearly such a correction is crucial in 
measuring the lensing signal.

The lower panel of Figure \ref{map.ps} shows the
R-mode of the mass aperture statistic. 
As this statistic uses a compensated filter, 
the scale beyond which the measured R-mode is consistent with zero
($5'$ on the plot) corresponds to an effective angular scale 
$\theta \simeq 1'$. This
places an upper limit on measured shear correlations due to 
the intrinsic alignment of galaxies, given the redshift distribution of
the sources. The vanishing of $\langle M_\perp^2\rangle$ for 
effective angular scales larger
than $1'$ strongly supports our conclusion that the level
of residual systematics is low: this is a very hard test to 
pass, as it means that the signal is produced by a pure scalar field, 
which need not be the case for systematics. We checked that $M_\perp^2$
is Gaussian distributed with a zero average all over the survey, as
what we would expect from a pure noise realisation. For 
scales below $5'$ on the plot, the
$R$-mode is not consistent with zero at the 2-$\sigma$ level. Since the
cross-correlation
$\langle \gamma_t\gamma_r\rangle_\theta$ is consistent with
zero at this scale, the
source of the $R$-mode is probably not a residual systematic.
It might be due to the effect of intrinsic alignments (\cite{CNPT00b}),
but it is difficult to be sure without further tests. 

The error bars shown in Figures \ref{tophat.ps} to \ref{crosscorrel.ps} 
are calculated from a measurement
of the different statistics in $200$ realizations of the data set,
with randomized orientations of the galaxies. We measured the 
sample variance from ray-tracing simulations (\cite{JSW00}) and find
that it is smaller than $20\%$ of the noise error bars shown here 
(see \cite{VW99} where the sample variance has been calculated for surveys with
similar geometry), therefore we have not included it in our figures.
Figure \ref{cosmicvar.ps} shows an estimate of the
sample variance for the r.m.s. shear using a compact $6.5$ sq. degree
ray-tracing simulation (\cite{JSW00}). This figure shows that
the sample variance is about order of magnitude smaller than the 
signal for the range of scales of interest. Hence our errors are not 
dominated by sample variance, as was the case in the first detections 
of cosmic shear. As the probed
angular scales approach the size of the fields (which is $\sim 30'$
with the CFH12K camera) the sample variance becomes larger. 
This could be responsible for the small fluctuations in the measured
correlations in Figures \ref{gg.ps} and \ref{eteterer.ps} for scales larger
than $24'$.


\section{Cosmological constraints}

As noted elsewhere (e.g. \cite{B97,JS97}), the parameters that
dominate the 2-point shear statistics are the power
spectrum normalization $\sigma_8$ and the mean density $\Omega_0$.
We investigate below how the statistics measured
in Figures \ref{tophat.ps} to \ref{eteterer.ps} constrain these
parameters. Our parameter estimates below rely on some simplifying 
assumptions; a more detailed analysis over a wider space of parameters 
will be presented elsewhere. 

We assume that the data follow Gaussian statistics and neglect sample variance
since it is a very small contributor to the noise for 
our survey, as discussed above. We compute the likelihood function $\cal L$:

\begin{equation}
\Lc={1\over (2\pi)^{n/2} \left|\Sg\right|^{1/2}}~\exp{\left[-{1\over 2}\left({\rm \dg-\sg}
\right)^T \Sg^{-1}\left({\rm \dg-\sg}\right)\right]},
\label{likelihood}
\end{equation}
where $\dg$ and $\sg$ are the data and model vectors respectively, and
$\Sg:=\langle\left({\rm \dg-\sg}\right)^T\left({\rm \dg-\sg}\right)\rangle$
is the noise correlation matrix. $\Sg$ was computed for the 
different statistics
from 200 random realizations of the survey, therefore effects associated
with the survey geometry are included in our noise matrix.
The model $\sg$ was computed for a grid of cosmological models which covers
$\Omega_0\in [0,1]$ and $\sigma_8\in [0.2,1.8]$ with a zero cosmological
constant. The prior is chosen to be flat
over this grid, and zero outside. We also fixed $\Gamma=0.21$ and used the redshift
distribution of Eq.(\ref{zsource}). We discuss below the impact of this
choice of priors.

Figure \ref{map.ps} (bottom panel) shows that for effective scales 
smaller than $1'$ there is a non-vanishing R-mode which could come either
from a residual systematic, or from an intrinsic alignment effect. Therefore
it is safer to exclude this part from the likelihood calculation. Thus
for the top-hat variance, we excluded the point at $1'$, for the correlation
functions the points below $2'$, and for the $M_{\rm ap}$ statistic the points
below $5'$. For the correlation function, we also excluded the points at
scales larger than $20'$ because of the small fluctuations in the measured
correlations. The constraints
on the cosmological parameters are not significantly affected whether
these large scale points are excluded or not. 

\begin{figure}
\centerline{
\psfig{figure=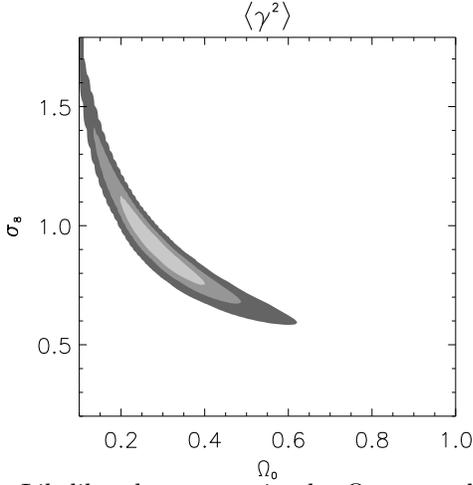,height=6cm}
}
\caption{\label{omega_sigma_var.ps} Likelihood contours 
in the $\Omega_0 -\sigma_8$ plane from the 
top-hat smoothed variance $\langle\gamma^2\rangle$
shown in Figure \ref{tophat.ps}. The first point in Figure \ref{tophat.ps}
was not included in the likelihood calculation to avoid the small
scale systematic shown in Figure \ref{map.ps} (bottom panel).
The cosmological models have $\Lambda=0$,
with a CDM-type power spectrum and $\Gamma=0.21$. The redshift of the
sources is given by Eq.(\ref{zsource}). with $(z_0,\alpha,\beta)=(0.8,2,1.5)$.
The confidence levels are $(0.68, 0.95, 0.999)$.
}
\end{figure}

\begin{figure}
\centerline{
\psfig{figure=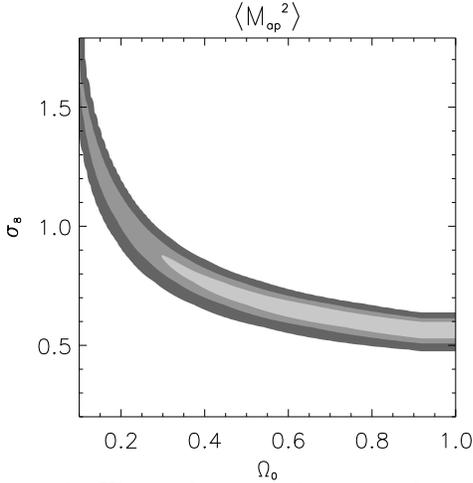,height=6cm}
}
\caption{\label{omega_sigma_map.ps} As in Figure \ref{omega_sigma_var.ps},
but using the $M_{\rm ap}$ statistic of (Figure \ref{map.ps},
top panel)
instead of the top-hat variance. The first five points in Figure \ref{map.ps}
were not included in the likelihood calculation in order to avoid the 
small scale systematic shown in Figure \ref{map.ps} (bottom panel).
}
\end{figure}

\begin{figure}
\centerline{
\psfig{figure=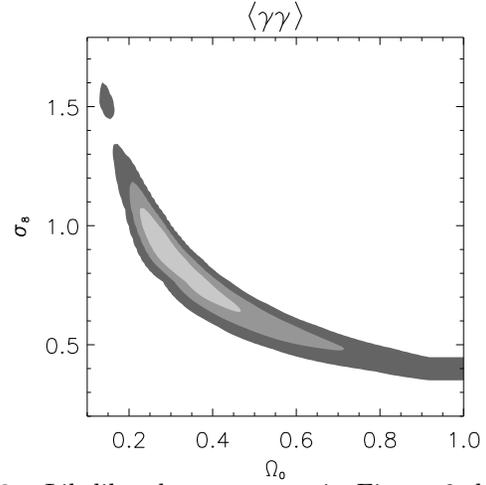,height=6cm}
}
\caption{\label{omega_sigma_gg.ps} Likelihood contours 
as in Figure \ref{omega_sigma_var.ps},
but using the shear correlation function $\langle\gamma\gamma\rangle_\theta$
(Figure \ref{gg.ps})
instead of the top-hat variance. The first two points and scales larger
than $20'$ in Figure \ref{gg.ps}
were not included in the likelihood calculation to avoid the
contribution from the small
scale systematic shown in Figure \ref{map.ps} (bottom panel).
}
\end{figure}

\begin{figure}
\centerline{
\psfig{figure=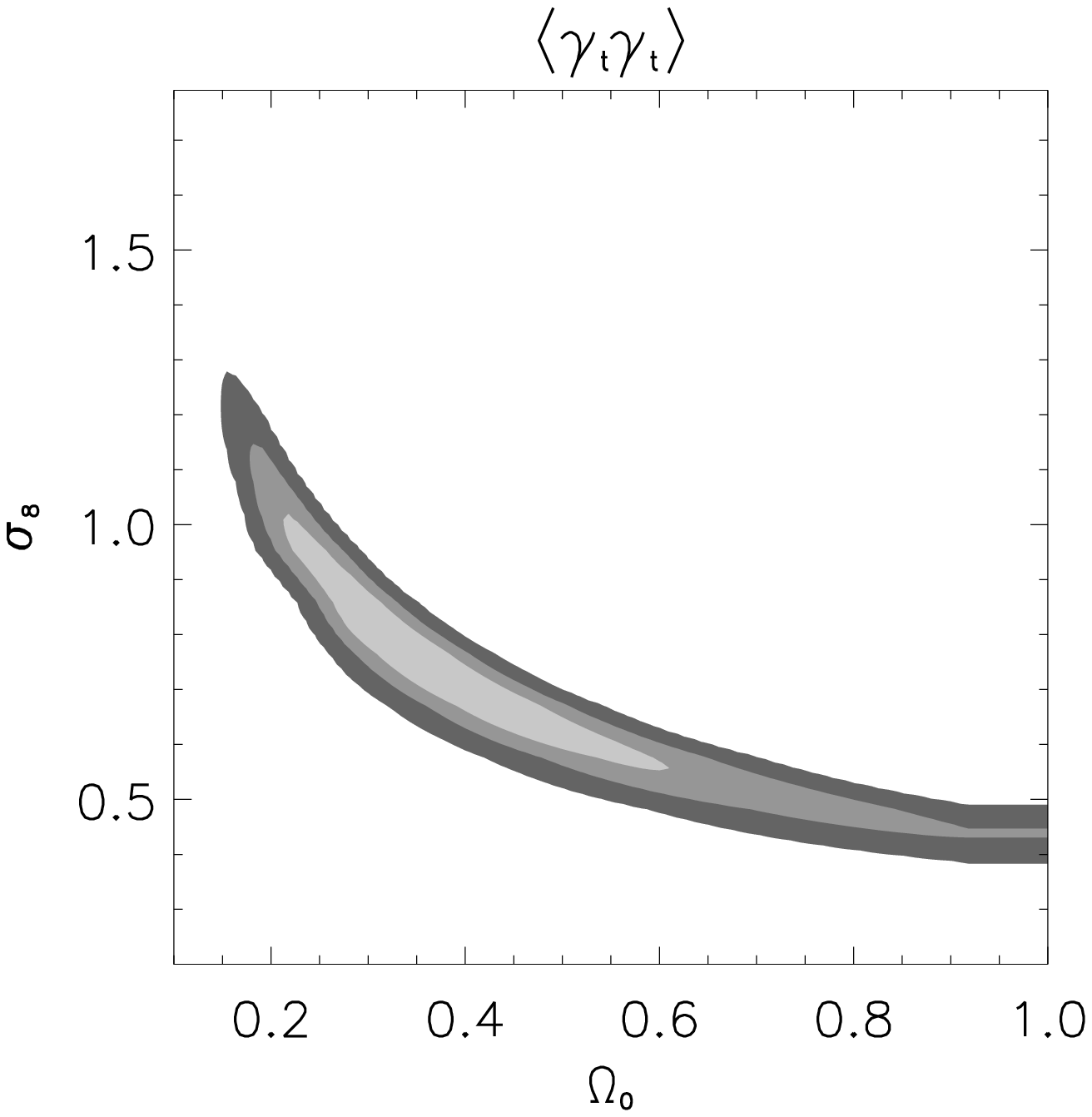,height=6cm}
}
\caption{\label{omega_sigma_etet.ps} As in Figure \ref{omega_sigma_var.ps},
but using the tangential shear correlation function
$\langle\gamma_t\gamma_t\rangle_\theta$ (Figure \ref{eteterer.ps})
instead of the top-hat variance. The first two points and scales larger
than $20'$ in Figure \ref{eteterer.ps}
were not included in the likelihood calculation in order to avoid the
contribution from the small
scale systematic shown in Figure \ref{map.ps} (bottom panel).
}
\end{figure}

\begin{figure}
\centerline{
\psfig{figure=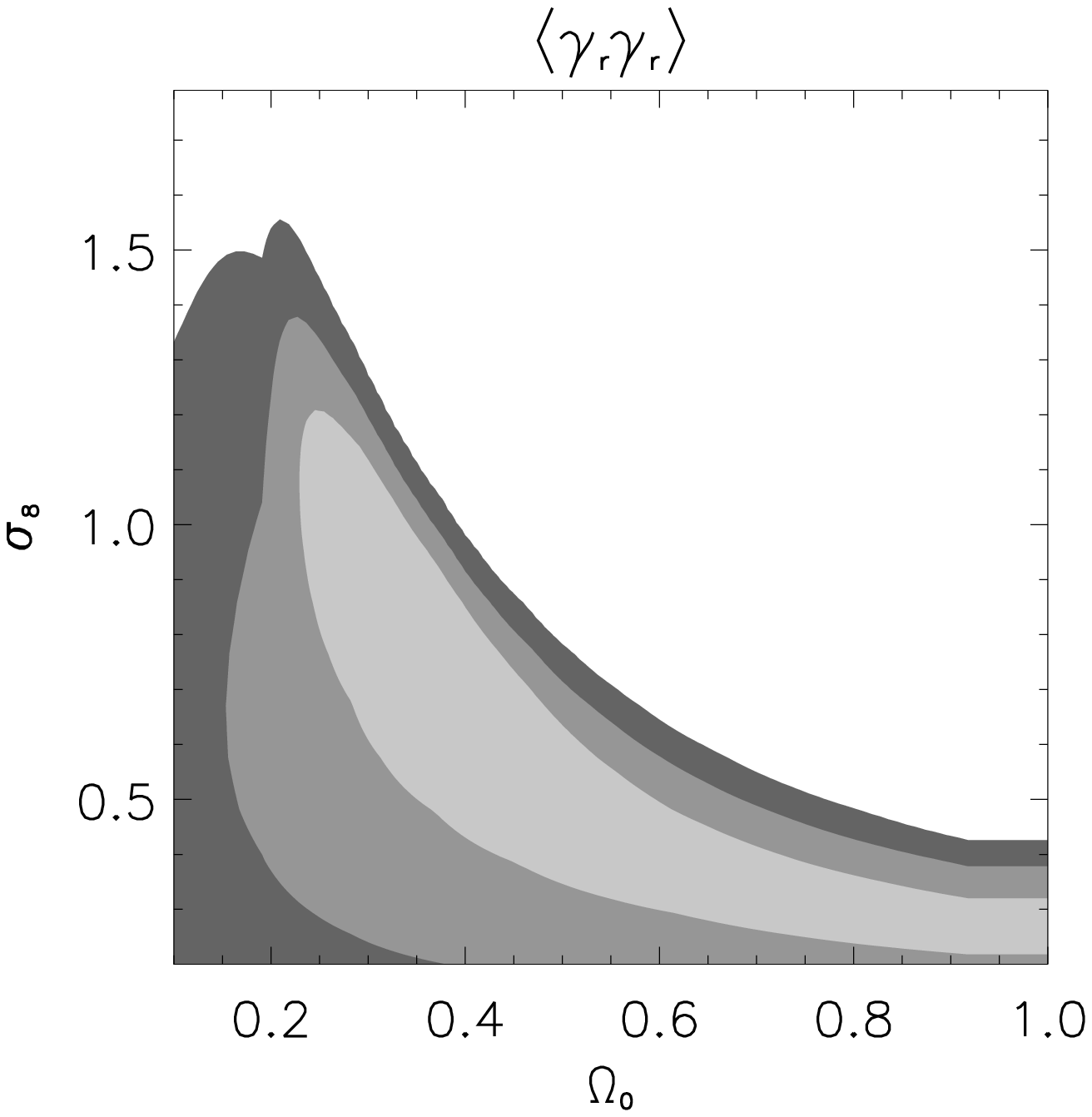,height=6cm}
}
\caption{\label{omega_sigma_erer.ps} As in Figure \ref{omega_sigma_var.ps},
but using the radial shear correlation function
$\langle\gamma_r\gamma_r\rangle_\theta$ (results in Figure \ref{eteterer.ps})
instead of the top-hat variance. The first two points and scales larger
than $20'$ in Figure \ref{eteterer.ps}
were not included in the likelihood calculation in order to avoid the
contribution from the small
scale systematic shown in Figure \ref{map.ps} (bottom panel).
}
\end{figure}

\begin{figure}
\centerline{
\psfig{figure=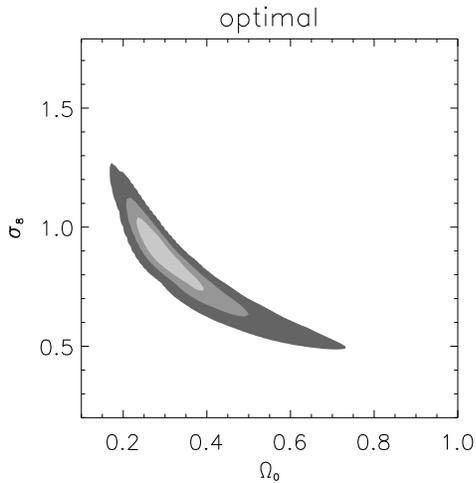,height=6cm}}
\caption{\label{omega_sigma_optimal.ps} Likelihood contours as in 
Figure \ref{omega_sigma_gg.ps},
but all the points in Figure \ref{gg.ps} on scales smaller than 
$20'$ were used. In order to
account for the small scale systematic shown in Figure \ref{map.ps} 
(bottom panel)
the error bars on the first two points were increased to include 
the systematic amplitude. 
}
\end{figure}

Figures \ref{omega_sigma_var.ps} to \ref{omega_sigma_erer.ps} show the
$(\Omega_0,\sigma_8)$ constraints for each of the statistics shown in
Figures \ref{tophat.ps} to \ref{eteterer.ps}. The contours show the $99.9\%$,
$95.0\%$ and $68.0\%$ confidence levels. The agreement between the
contours is excellent, though the $M_{\rm ap}$ statistic and the radial
correlation function do not give as tight constraints as the other
statistics. The correlation function measurements below 
$2'$ may be considered by using error bars that include a possible 
systematic bias: this is equivalent to adding a systematic
covariance matrix $\Sg^{sys}$ to the noise covariance $\Sg$ matrix
in Eq.(\ref{likelihood}). The new contours
computed with the enlarged error
bars\footnote{The enlarged error bars were computed from the estimation of our
$B$-mode analysis which will be presented elsewhere} are shown
in Figure \ref{omega_sigma_optimal.ps}. 
The maximum of the likelihood in the variance and correlation function
likelihood plots is at $\sigma_8\simeq 0.9$ and $\Omega_0\simeq 0.3$. 
Note that compared to a similar plot in \cite{Ma01}
(Figure 8), here the contours are narrower, and are obtained 
from a homogeneous data set. Moreover, the degeneracy
between $\Omega_0$ and $\sigma_8$ is broken. 

The partial breaking of degeneracy between 
$\Omega_0$ and $\sigma_8$
was expected from the fully non-linear calculation of shear correlations
(\cite{JS97}). In the non-linear regime the dependence of the 
$2$-points statistics on $\Omega_0$ and $\sigma_8$ becomes 
  sensitive to angular scale.
 For example, as shown in \cite{JS97},  
the shear r.m.s. measures $\sigma_8~\Omega_0^{0.5}$ on scale between $2'-5'$,
and $\sigma_8~\Omega_0^{0.8}$ on scales $\gsim 10'$. Therefore a low $\Omega_0$
universe should see a net decrease of shear power at large scale compared to
a $\Omega_0=1$ universe (for a given shape of the power spectrum), 
as evident in Figure \ref{tophat.ps}. Note that the aperture mass
$M_{\rm ap}$ is still degenerate with $\Omega_0$ and $\sigma_8$
(Figure \ref{omega_sigma_map.ps}) because it probes effective scales up to
$\sim 2.6'$ only, which is not enough to break the degeneracy.

It seems that the aperture mass (Figure \ref{omega_sigma_map.ps}) gives
a slightly larger $\sigma_8$
for a large $\Omega_0$ compared to the other statistics, while they all
agree for $\Omega_0 < 0.7$. This could be an indication
for a low $\Omega_0$ Universe, however in practice,
the probability contours for the different statistics cannot be
combined in a straightforward way because they are largely redundant.
The best strategy here is to concentrate on one particular
statistic.
We expect the best constraints from the 
shear correlation function (since it contains all the information
by definition), and therefore base our parameter estimates 
on the likelihood contours obtained from it. 
The contours in the $\sigma_8-\Omega_0$ plane in 
Figure \ref{omega_sigma_optimal.ps} closely follow 
the curve $\sigma_8\propto \Omega_0^{0.6}$. This allows us
to obtain the following measurement of $\sigma_8~\Omega_0^{0.6}$
(from this figure alone):
\begin{equation}
\sigma_8~\Omega_0^{0.6}\ =\ 0.43^{+0.04(0.06)}_{-0.05(0.07)}, 
\label{sigma8Omega}
\end{equation}
where the uncertainties correspond to the $95\%$ ($99.9\%$) 
confidence levels. The result in equation (\ref{sigma8Omega})
is fairly robust against different values of $\Gamma$.

If we fix $\Gamma=0.21$, we can constrain the two parameters separately; 
we get, at the $95\%$ confidence level:
$0.2< \Omega_0 <0.5$ and $0.6< \sigma_8 <1.1$ for open models
and $\sigma_8>0.65$ and $\Omega_0<0.4$ for flat ($\Lambda$-CDM) models.
This result however is sensitive to
the prior choosen for $\Gamma$. In particular, if we use the
relation $\Gamma=\Omega_0 h$ for a cold dark matter model, 
then some extreme combinations of $\sigma_8$, $\Omega_0$ and
$\Gamma$ cannot be ruled out from lensing alone. The degeneracy between
$\Omega_0$ and $\sigma_8$ is broken only if we take $\Gamma$ to lie in
a {\it reasonable} interval. Such interval can be motivated by
galaxy surveys for instance, which give
$0.19<\Gamma<0.37$ at $68\%$ confidence level for the APM
(\cite{EZ01}). Therefore the separate constraints on $\Omega_0$ and 
 $\sigma_8$ given above require some prior assumptions and must be taken
with precaution, while 
the constraint on $\sigma_8~\Omega_0^{0.6}$ is much more
robust. The redshift distribution of the sources is likely to be 
the main source of uncertainty in our estimate of 
equation (\ref{sigma8Omega}); a rough guide is given by the scaling 
$\sigma_8~\Omega_0^{0.6} \propto z_0^{-0.5}$ (\cite{JS97}).  
A more detailed analysis of parameter estimation is left for 
a later study.

\section{Small scale signal}

\begin{figure}
\centerline{
\psfig{figure=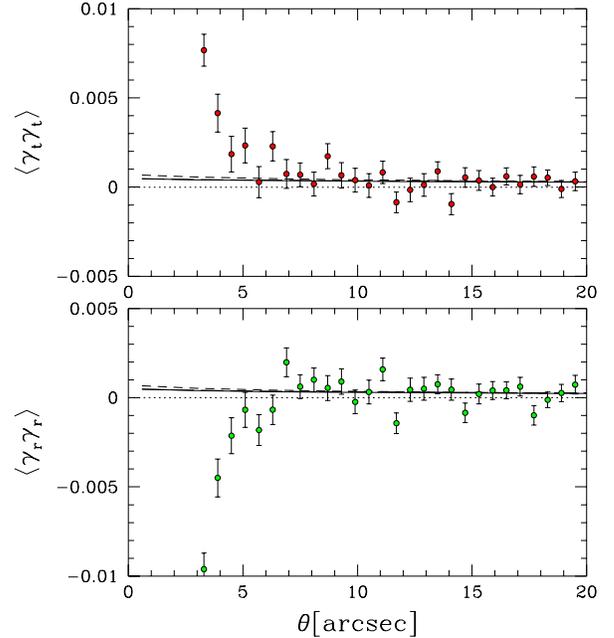,height=9cm}}
\caption{\label{smallscale_eteterer.ps} Tangential (top panel) and
radial (bottom panel) shear correlation
functions $\langle\gamma_t\gamma_t\rangle_\theta$ and
$\langle\gamma_r\gamma_r\rangle_\theta$ down to $3''$. The solid, long-dashed
(hidden by the solid line) and
short-dashed lines are predictions from the same models as in Figure
\ref{tophat.ps}.
}
\end{figure}

\begin{figure}
\centerline{
\psfig{figure=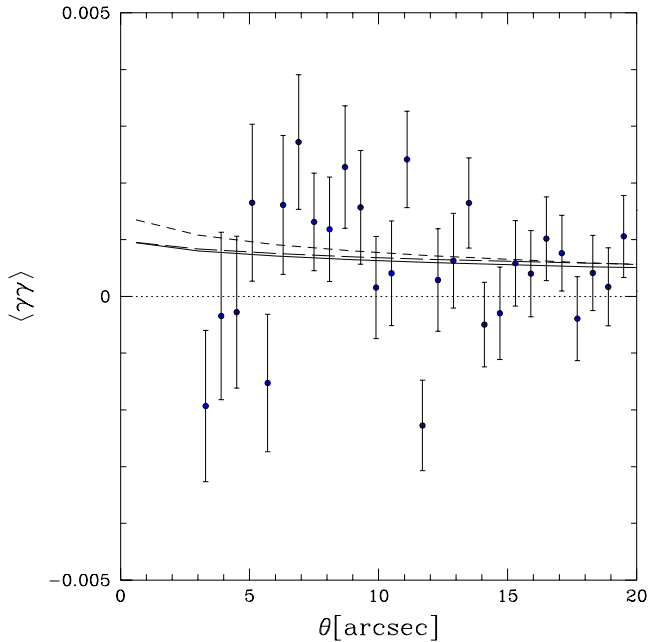,height=9cm}}
\caption{\label{smallscale_gg.ps} Same as Figure \ref{smallscale_eteterer.ps}
but for the shear correlation function
$\langle\gamma\gamma\rangle_\theta$.
}
\end{figure}

Our correlation function measurements extend to much smaller scales
than shown in the figures above. The limit is set only by the fact
that we reject one member of all pairs closer than $3$ arcsec. 
Figures \ref{smallscale_eteterer.ps} and \ref{smallscale_gg.ps}
show the tangential, radial and total shear correlation functions. The pair
separation bins are much smaller than in Figures 
  \ref{gg.ps} and \ref{eteterer.ps}, which explains why the error bars are larger. 
Even at the smallest scales, the shear
correlation function $\langle\gamma\gamma\rangle_\theta$ is consistent
with the model predictions. 

The surprising result for the small scale correlations is the behavior 
of the tangential and radial shear correlation functions: at scales 
smaller than $5''$ we
find an increased amplitude for $\langle\gamma_t\gamma_t\rangle_\theta$,
and a ${\it negative}$ $\langle\gamma_r\gamma_r\rangle_\theta$. 
 Though surprinsing, a negative $\langle\gamma_r\gamma_r\rangle_\theta$ is not unphysical: for
instance in \cite{K92} (Table 1) a shallow mass power
spectrum ($n>-1$) implies such an effect. In terms of halo mass profile, 
it corresponds to a projected profile steeper than $-1.5$.
 However, regardless of the nature of this signal, it is
important to note that this is a
very small scale effect which has no effect on the statistics discussed 
in preceding sections. The contribution of the increased signal from 
$\langle\gamma_t\gamma_t\rangle_\theta$
to the variance at $1'$ is less than $1\%$; moreover since
$\langle\gamma\gamma\rangle_\theta$ is not affected at all, 
the variance is also
unaffected. As an explicit test, 
we checked that by removing one member of the pairs closer than
$7''$ the measured signal in
Figures \ref{tophat.ps},\ref{map.ps},\ref{gg.ps},\ref{eteterer.ps} is
unchanged. In a similar cosmic shear analysis using the
Red-sequence Cluster Survey
(\cite{G00}) another group finds a similar small scale behavior, though
at lower statistical significance (H. Hoekstra, {\it private communication}).

The cross-correlation $\langle\gamma_t\gamma_r\rangle_\theta$ vanishes down
to $3''$, therefore no obvious systematic is responsible for this
effect. The effect is unlikely to be 
caused by overlapping isophotes, or
close neighbors effects because $\langle\gamma_t\rangle_\theta^2 <<
\langle\gamma_t\gamma_t\rangle_\theta$: if it were a close neighbor alignment
we would expect that $\langle\gamma_t\rangle_\theta$ (the average
tangential ellipticity
for all the pair members in each pair separation bin $\theta$) carries all of
the signal, which is not the case. In fact we find $\langle\gamma_t\rangle_\theta^2 \sim 0.2 \langle\gamma_t\gamma_t\rangle_\theta$, which means that
close neighbor effect can hardly exceed $20\%$ of the small scale
signal.

A forthcoming paper using the same data set will be devoted to the measurement
of $E$ and $B$ modes (as defined in \cite{CNPT00a}), and we will study this
small scale signal in more detail. At this stage of the analysis we
cannot exclude a possible residual systematic.
However, a preliminary analysis shows that the $B$ mode down to $3''$ is much
smaller than the $E$ mode, which is hard to have if the signal comes
from residual systematics.

\section{Conclusion}

Using $6.5$ sq. deg. of the VIRMOS survey in progress at the CFHT, we were
able to measure various $2$-points correlation 
statistics of cosmic shear. The top-hat variance, the aperture mass statistic 
and different shear correlation
functions gave consistent results over a wide range of scales. Further
tests of the lensing origin of the signal exploiting the scalar nature
of the gravitational potential were also convincingly verified. 
We demonstrated that the level of systematics, in particular the intrinsic
alignment of galaxies, is likely to be small, and does not contribute 
to the signal for scales larger than $1'$.
We believe that these results demonstrate the significance of our detection
of shear correlations due to gravitational lensing. The quality of the
data and the adequate size of our survey allow us to constrain 
cosmological models of the large-scale distribution of dark matter 
in the universe.  

We have obtained tight constraints on the cosmological
parameters $\Omega_0$ and $\sigma_8$. These results suggest that
high precision measurements can be made 
with larger surveys on a much larger set of
cosmological parameters. The final
stage of the VIRMOS survey is to accomplish $16$ sq. deg. in patches
of $4$ sq. deg., $4$ colors each, thus allowing the possibility to use
the photometric redshifts of the galaxies. The use of photometric redshift
will not only improve the scientific interpretation of cosmic shear (e.g.
doing tomography as in \cite{W99}) but will be useful to measure
the intrinsic alignment
itself (which can be used to constrain galaxy formation models for instance).

The constraints obtained so far are within a framework of
structure formation through gravitational instability with Gaussian
initial conditions and Cold Dark Matter.
As the amount of observations increases and the
measurement quality improves,
the first hints of the shape of the power spectrum will be soon available.
It opens new means of
really testing the formation mechanisms of the large-scale structure
and the cosmological parameters beyond the standard model (\cite{UB00}).

Over the last two years, we have seen the
transition from the detection of the weak lensing signal to the first
measurements of cosmological parameters from it. The agreement
between theoretical predictions and observational results with
such a high precision indicates that the measurement of 
cosmic shear statistics is becoming a mature cosmological tool.  
Many surveys are under way or scheduled for the next 5 years.
They will use larger panoramic cameras than the CFH12K, and will
cover solid angles 10 to 100 times wider than this work. 
The results of this work give us confidence that
cosmic shear statistics will provide valuable measurements of
cosmological parameters, probe the biasing of mass/light, 
produce maps of the dark matter distribution and reconstruct its
power spectrum. 

{\acknowledgements 
We are grateful to St\'ephane Colombi, Ue-Li Pen, Dmitri Pogosyan, Simon Prunet,
Istvan Szapudi and Simon White for useful discussions
related to statistics. We thank Henk Hoekstra for sharing his
results prior to publication.
This work was supported by the TMR Network ``Gravitational Lensing: New
Constraints on
Cosmology and the Distribution of Dark Matter'' of the EC under contract
No. ERBFMRX-CT97-0172, and a PROCOPE grant No. 9723878 by the DAAD and
the A.P.A.P.E. We thank the TERAPIX data center for providing its facilities
for the data reduction of CFH12K images.  
}

\end{document}